\documentclass[12pt]{article}
\usepackage{url}
\usepackage{epsfig}
\usepackage{amsmath}
\usepackage{amssymb}
\usepackage{amsthm}
\usepackage{latexsym}
\usepackage{graphicx}
\usepackage{subfigure}
\usepackage{fancyheadings}
\begin{document}

\title{
Can Random Coin Flips Speed Up a Computer?
}
\author{David Zuckerman\thanks{
Email: \texttt{diz@cs.utexas.edu}.
Supported in part by NSF Grants CCF-0634811 and CCF-0916160.}\\
University of Texas at Austin
}

\maketitle

\section{The Power of Randomness}

Randomness is everywhere.
Do you like to cook?
Imagine you're sauteeing some onions.
You'd like each slice of onion to be equally browned on both sides.
However, you don't measure
this for each piece.
Instead, you stir randomly, and by the ``law of large numbers,''
this occurs naturally.

The law of large numbers says that if we flip a fair coin many times,
about half the time it will come up heads and half tails.
Of course, we expect some error:  heads may appear 53\% of the time, for example.
We call this 3\% difference from the expected 50\% the error.
The error gets smaller the more times we flip the coin, becoming
negligible for a large number of flips.
 
We all pay attention to polls before elections.
Surprisingly, a modest sample of 400 random people can predict percentages of the whole
United States to within 5\% (with 95\% certainty).\footnote{
Real polls have additional error because the sample may not be random,
but that is a different topic.}
This idea is called random sampling.

Random sampling relies on the law of large numbers.
To see this, suppose 57\% of the population supports Obama.
Asking a random person whether they support Obama is the same as flipping a biased coin,
which comes up ``yes'' 57\% of the time and ``no'' 43\% of the
time.  By the law of large numbers, the more people we ask, the smaller the error.
Note that this argument doesn't depend on the size of the population; the error would be
the same regardless of the population of the United States.

Want to make money?
It's commonly assumed
that the price of a stock follows a
``random walk."  That is, each day the price change corresponds to a random
coin flip, with a slight bias upwards.
(If there were no bias upwards, there would be no incentive to buy stock.)
If you invest \$100 in stocks, your daily return can be roughly approximated by a coin flip,
where half the time you make $\$1.20$ and half the time you lose $\$1.15$.
This might make stocks seem way too risky, given the minimal gain.
Yet over time, numerous coin flips cancel each other out
and the bias upwards starts to grow.  Over many years, the bias is very
apparent.

The law of large numbers is the real reason for diversifying your investments.
If you own many different securities whose prices move independently
of each other (or almost independently), you reduce your risk.

Investing differs from our earlier examples of cooking and polling,
in that the randomness is modeling reality, and is not something we control.
In cooking and polling, we \emph{choose} to add randomness to achieve
our aims.

That's what happens in computing.  There are many
situations where we choose to add randomness, i.e., random coin flips, to compute quicker.
In this essay, we describe several such situations.
Some applications rely on the law of large numbers in the guise
of random sampling, while many do not.

After describing the benefits of randomness, we
explore how to get good randomness.
Finally, we briefly discuss whether the benefits of randomness are real,
or whether there are equally-good methods which avoid adding randomness.
 
\cfoot{\thepage}
\rfoot{}

\subsection{Uses of random sampling}

Before proceeding further, we define random coin flips and random numbers.
A random coin flip means that with probability 1/2 we get heads, which we'll call 1,
and with probability 1/2 we get tails, which we'll call 0.
Similarly, by a random number between 1 and 100, we mean that each number
between 1 and 100 has probability 1/100 of occurring.

More generally, we may need to generate several coin flips or random numbers.
By a random sequence of coin flips, we mean that the coin flips
are random and ``independent":  no matter what the values of all other coin flips, a particular
coin still has probability 1/2 of being heads and 1/2 of being tails.

We begin with two
applications of random sampling in computing.
For our first example, suppose we have several computers available to
perform many tasks.
The tasks require different amounts of time, which are unknown in advance.
How can we distribute the tasks to the different computers so that the loads of the different computers
will be roughly equal?  You guessed it:  distribute them randomly.
By balancing the loads, the computers will finish quicker.

Our second example involves computing the area of a complicated object.
The only understanding we have about this object is that if we are given a point,
we can tell whether or not it is inside this object.
We can estimate the area of this object by imagining it
being inside a simpler object,
such as a square.

\bigskip
\centerline{\epsfig{file=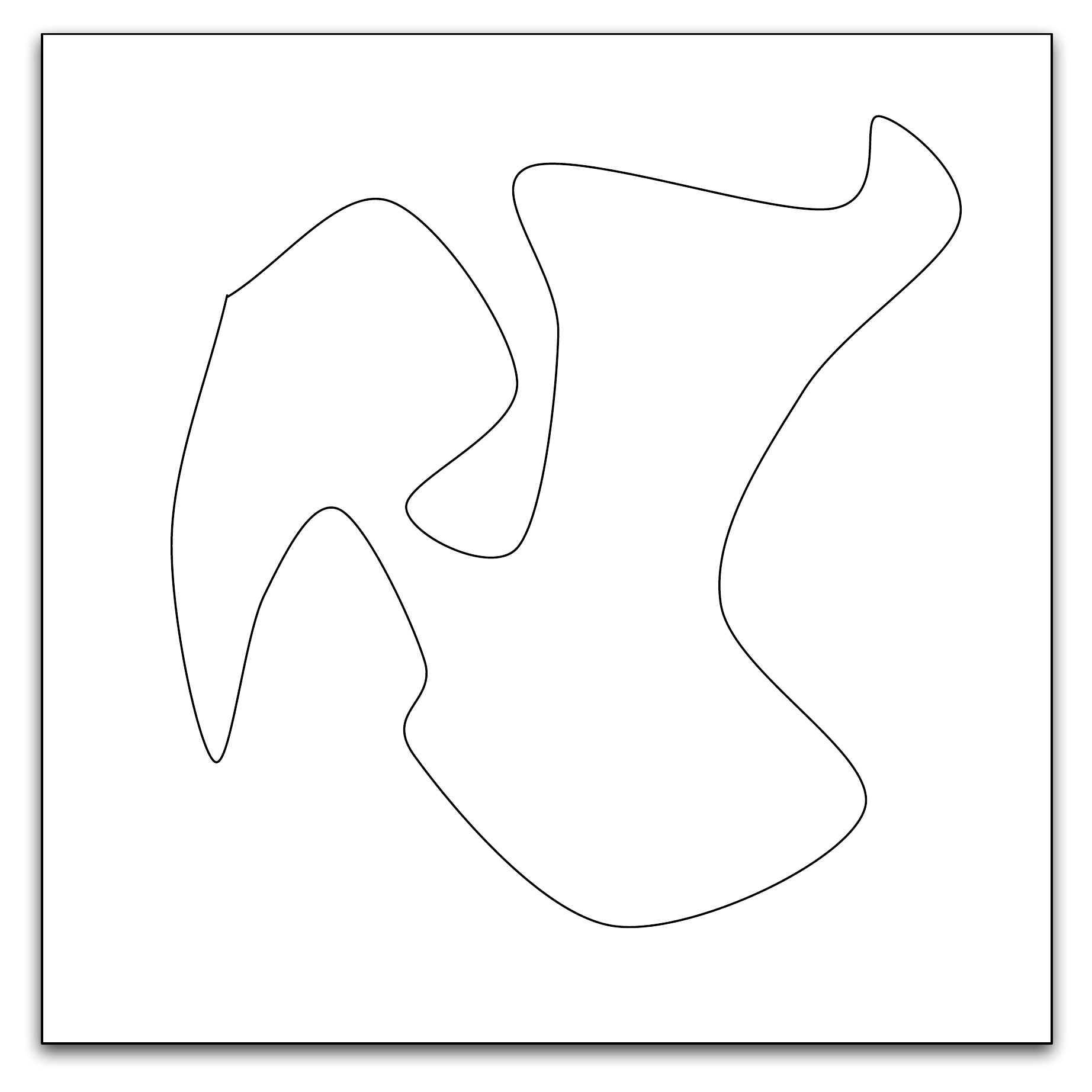,height=1.3in}}

We can then sample random points from the square,
and count how many lie inside our complicated object.
If 23\% of the sampled points lie in the object, we can estimate
the area of the object to be 23\% of the area of the square.
Since it's easy to calculate the area of the square, we have our estimate.
This method extends naturally to approximating volumes.

These are examples of randomized \emph{algorithms}.  An algorithm is simply a recipe
to compute something.
For example, the simple grade-school recipe for adding numbers, using carries, is
an algorithm.  This addition algorithm has two inputs, the numbers to be added, and produces
an output, the sum of the two numbers.
Randomized algorithms typically involve a small probability of error.
Despite this, they are often very useful because of their simplicity and speed.

Randomized approximation algorithms have two types of error.
One is the error in the quality of the approximation.
We don't expect our algorithm to exactly compute the area of our complicated object,
but instead to approximate it, say within~1\%.

The other type of error occurs when, for example, we take enough samples
to be highly confident that our estimate is within 1\% of the truth, but we get unlucky and
our estimate is not that accurate.
For example, we could get very unlucky in our area estimation and all of our sampled points lie inside
the object, causing us to drastically overestimate the area.
This type of error is more problematic, but there are ways to guarantee that the probability of
this error is so small as to be negligible.
For example, we might ensure that the probability of this error is less than one in a billion.
This differs from other areas of science, where an error as high as 5\% may be acceptable.

\subsection{Other uses of randomness}

There are other types of randomized algorithms which don't rely on random sampling.
Probably the most famous randomized algorithm tests whether a number is prime.
(Recall that a number is prime if the only numbers dividing it evenly are 1 and itself.)
A number being prime seems so deterministic; how could randomness help?
Yet in the late 1970's, researchers devised a fast algorithm to solve this problem using randomness.
The algorithm is not to pick a random number and divide it into the input number, but
rather something more complicated which we won't discuss.

Instead, we give the following example,
which while less important,
illustrates an interesting technique.  If you find it too technical, simply skip to the
next section on Cryptography.

The problem is to test for equality of expressions.
For instance, we may test whether
\[ 12^{1,000,001} + 7^{442} = 143^{500,001} + 197. \]
As a puzzle, see if you can find a simple way to verify that, in fact, in this example
the two expressions are not equal.
You don't need a calculator or any paper to solve it.

Note that each side represents a number over a million digits long.
Multiplying out each side and comparing a million digits would be quite inefficient.

The key to solving the puzzle is parity: whether a number is odd or even.
The left side is the sum of an even number plus an odd number, and so is odd.
The right side is the sum of two odd numbers, and so is even.
Therefore, they can't be equal.

Is this just a cheap trick that works for this example?  At first glance,
yes.  If both sides were even (or both odd), that gives us no indication
that the two sides are actually equal.
On the other hand, an extension of this idea can work in general.
Note that odd and even is just the remainder when divided by 2.
Why not try the remainder when divided by 3,4, or 5?

A randomized algorithm to solve the general problem proceeds as follows.
Pick a random number $r$ of a suitable size.  In the above example,
$r$ should be a random number between 1 and 50,000,000.  
Compute the remainder of both sides when divided by $r$.
This doesn't require multiplying out both sides; in fact, it turns out to be efficient for a computer.
The remainder is guaranteed to be less than 50,000,000, which might seem big, but it is only 8 digits.
That's much shorter than a million digits.

If the two sides are equal, then of course the remainders when divided by any $r$
will also be equal.  One can show that if the two sides are unequal,
then there are few choices of $r$ which will cause the remainders to be equal.
In other words, the probability of getting an error by choosing such a bad $r$ is small.

In fact, we can make the error tiny.  Suppose the original
error is, say, 10\%, i.e., $1/10$.  Let's repeat the algorithm 10 times.  Now the error becomes
$(1/10)^{10}$, which is 1 in 10 trillion.
That should suffice for any practical purpose.
If we are really paranoid, we may repeat it, say, 100 times.
Then the error is less than 1 divided by the number of particles
in the universe.

There are many other randomized algorithms, most of which are more
sophisticated than this example.  As in this example, in most cases
one would not immediately expect randomness to be useful.  Yet randomness often
provides the fastest or simplest way to solve a problem.

Of course, randomness is also often necessary to model a problem.
Randomness is an essential ingredient for computer simulations
of complex phenomena,
such as the economy, the weather, and the interaction of molecules.
Sometimes the randomness is inherent in the process being simulated --
the laws of quantum mechanics involve randomness -- but typically
the randomness is a way to model complexity.

\subsection{Cryptography}

Besides being used in randomized algorithms, randomness is necessary for computer security.
Imagine you're using wireless on your laptop in a cafe.
You're buying something from a website.
It's possible for someone sitting at another table with their laptop to record all
the information sent and received by your computer.
If your computer and the website had agreed on a way to communicate secretly beforehand,
it seems believable that you could communicate securely.
Yet what happens if you never visited this website before?
How could you possibly communicate secretly?

At first, security seems impossible.  If the two honest computers
don't start with any shared secret information, and an adversary sees all
communication between the computers, it seems impossible for the
honest computers to convey any secret information.  And it is
impossible in some sense:  the adversary will have all the information to
compute any secret conveyed.

Yet public-key cryptography is a way to achieve this seemingly impossible task.
Somehow, although the adversary has the necessary information, it will take too
long to compute the secret from this information.
It works as follows.  The website can publish a way
to encrypt messages to it.  Your computer can publish an encrypted
message.  Yet only the website can decrypt your message.  Even
though the adversary knows the method of encryption and the encrypted
message, he still cannot decrypt the message in a reasonable amount of time.

The underlying building block for this is a ``one-way function.''
A one-way function is easy to compute but hard to invert.
To illustrate, think of a telephone book.
Given a name, it is easy to look up the corresponding phone number.
However, given a phone number, it is hard to look up the corresponding name.

As a more mathematical example, it is easy to multiply two prime numbers.
It doesn't take long to multiply two 4-digit numbers by hand.
However, it appears very hard,
given the product of two primes, to find the two primes (i.e., to factor the number).
You may recall the elementary algorithm for factoring a number:
simply divide successive prime numbers $2, 3, 5, 7,\ldots$ into the number.
While this isn't slow for small numbers, it is very inefficient for large numbers,
even for a computer.
Just think how long it would take to factor an 8-digit number by hand.

The idea behind the important public-key cryptosystem called RSA
is that the website chooses two primes
randomly (finally, we see where randomness comes in).
The encryption scheme depends only on the product of the two primes,
but decrypting requires knowledge of the actual primes.  Therefore,
only the website can decrypt.  RSA is a scheme that
achieves this.
Since RSA requires undergraduate-level knowledge of number theory, the branch of mathematics
that deals with prime numbers and related concepts, we won't explain it further.

\section{Getting Good Randomness}

Wait a minute.  We have these great uses of randomness, but do computers have access to
randomness?
In fact, computers can
get some randomness by looking at the exact time of a clock.
For example, if the exact time is 2:49.389917, the computer could use the last four digits,
9917, as a random number.
However, computers can't get too many digits like this.
If it starts looking at the clock in a regular
pattern, say every .001322 seconds, then consecutive 4-digit ``random'' numbers will differ
by 1322, and they won't be random.
It turns out to be very difficult for computers to obtain many random digits.

However,
to run our randomized algorithms we may need millions of random digits.
Computer scientists therefore try to convert a small number of random digits, called the
random seed, into a large number of ``random enough'' digits.
(We'll need to define ``random enough.'')
A pseudorandom generator (PRG) is an algorithm which does this.

\begin{center}
\begin{picture}(370,30)
\put(0,5){\makebox(80,20){99171322}}
\put(0,-13){\makebox(80,20){random seed}}
\put(80,15){\vector(1,0){30}}
\put(110,0){\framebox(45,30){PRG}}
\put(155,15){\vector(1,0){30}}
\put(185,5){\makebox(175,20){
2653589793238462643383279}}
\put(185,-13){\makebox(175,20){random enough?}}
\end{picture}
\end{center}

Intuitively, there is no way to create true extra random digits in a deterministic manner.
The output of a pseudorandom generator will \emph{not} be truly random.
So how could it be ``random enough''?

It helps to define a property that we want.  A first try might be that each digit in the output,
viewed in isolation,
should be random.
But then we could use a one digit random seed, and output the seed repeated over and over.
For example, if the random seed was 2, the PRG would output 222222222222222222222222222222.
Clearly, each output digit in isolation is random (all 10 digits have probability 1/10 of occurring),
but this is obviously not a good generator.

For a second try, let's require that each \emph{pair} of output digits be random.
For example, the third and 19th digit, viewed in isolation, will be random.
Surprisingly, it is possible to do this.

We illustrate this with a small example.
In fact, when thinking about math in general, it helps to start with small examples.
Our example will be that we have two good random coin flips, and we want to create three
random coin flips.  Instead of heads and tails, let's call the outcome of our flips 0 and 1.
That way we can sound technical and call them bits.

Our two random bits have four possible outcomes,
00, 01, 10, and 11, each occurring with probability 1/4.
If we had three random bits, there would be 8 possible outcomes,
000, 001, 010, 011, 100, 101, 110, and 111, each occurring with
probability 1/8.  There's no way to convert events which occur with probability
1/4 into events that occur with probability 1/8 without adding random coin flips.
Therefore, we can't output three truly random bits.

Since that's impossible, let's give ourselves the easier goal we alluded to earlier.
Instead of the three bits together being random, let's just ask if each pair can
appear random.
In other words, the first and second bits will appear random;
the first and third will appear random; and the second and third will appear random.
These bits are called \emph{pairwise independent}.
The following table shows a pairwise independent distribution on three bits.

\begin{center}
\begin{tabular}{|c|c|c|c|} \hline
$b_1$ & $b_2$ & $b_3$  & Probability  \\ \hline
0 & 0 & 0           & 1/4 \\ \hline
0 & 1 & 1           & 1/4 \\ \hline
1 & 0 & 1           & 1/4 \\ \hline
1 & 1 & 0           & 1/4 \\ \hline
\end{tabular}
\end{center}

Look down the first and second columns.  Note that each pair 00, 01, 10, and 11 appears,
each with equal probability.
Note this for the first and third columns, and for the second and third columns.

If the only requirement of the output is pairwise independence, then there are pseudorandom
generators outputting many bits with a very short seed.
For example, using just 20 random bits as seed, such a generator could output a million
pairwise independent bits.

But is pairwise independence random enough?  For some purposes it is.
Pairwise independent distributions follow the law of large numbers, so they are
useful for random sampling.  They also have a wide variety of other uses.

Pairwise independence isn't always useful.
For example, in cryptography much better random numbers are needed.
There is a very successful theory of pseudorandomness, but we won't elaborate further.

\subsection{Randomness Extraction}

Instead, we'll consider a different scenario:  suppose we have a lot of randomness, but its quality
is low.  Is this useful?
An example of low-quality randomness is the following type of probability distribution on the numbers
from 1 to 100.  Some of these numbers have probability 0, and others as high as 1/20.

Our first instinct is to convert this low-quality randomness into high-quality randomness.
Yet this turns out to be impossible in general.
Instead, it turns out that we can do this conversion with the help of a short random string.
The output won't be exactly random, but extremely close to random.
(This notion of almost-random is much more random than pseudorandom, but we won't
elaborate.)

\begin{picture}(400,150)(-100,0)
\put(-25,0){\framebox(200,20){almost random}}
\put(75,50){\vector(0,-1){30}}
\put(15,50){\framebox(120,40){{\Large Extractor}}}
\put(210,70){\vector(-1,0){75}}
\put(210,60){\framebox(60,20){random}}
\put(75,120){\vector(0,-1){30}}
\put(-40,120){\framebox(230,20){weakly random}}
\end{picture}

This is called a \emph{randomness extractor}, or simply an extractor.
Extractors are useful because it is easier to get a short random
string than a long one.
In fact, for some purposes, a short random string can be simulated
by trying all possibilities.  For example, sometimes we can replace
3 random digits by cycling over all $10^3=1000$ possible values of
these digits.  It may be that the factor of 1000 overhead is not prohibitive.

Surprisingly, this idea about cycling over all possibilities can produce a pseudorandom
generator, which doesn't seem to involve weak random sources.
Imagine we have an extractor which takes as input a 10-digit low-quality source and 3-digit
random string, and outputs an 8-digit almost-random string.
By cycling over all 1000 possibilities for the 3 digits, we can view this extractor as mapping
one 10-digit string to 1000 8-digit strings.
This mapping turns out to be a good pseudorandom generator for certain purposes.

For example, suppose we want to do a poll, and require a sample of 1000 adults in
the United States.  For simplicity, say there are 100 million people in the U.S.
Then we need a random 8-digit number to pick a person, and 1000 8-digit numbers
to do our poll.  That's 8000 random digits.  By using this extractor-based pseudorandom
generator, it suffices to pick just one 10-digit random number!

Similarly, suppose we wish to run the equality test discussed earlier, which
requires 8 random digits to get error .1.  To reduce the error to 1 in a billion,
we need to repeat it 9 times.  That's 72 random digits in total.  Instead, an extractor-based
pseudorandom generator
enables us to use just 18 random digits.

Another unexpected use of extractors is to cryptography.
We've seen that it's possible to communicate secretly without agreeing on a secret key
beforehand; nevertheless, it is more efficient if you have agreed on such a secret key.
Imagine that you're a spy, and before you leave the country, you and your handler agree on a secret,
randomly chosen 100 digit number to serve as the secret key.
Let's say you only need a 40 digit secret key to communicate secretly.
You store parts of the key in different places, worrying that no one place is safe.

Now let's say your worries are well-founded, and you believe a double-agent has seen
part of your key, but you don't know which part.
If the entire 100 digits are compromised, there is nothing you can do.
But suppose the agent has information about 50 digits.
We can model the uncertainty from the agent's point of view as randomness.
That is, from the agent's point of view, your 100 digits look like 50 fixed digits
(the known digits) and 50 random digits (the unknown digits).
This is a type of weak random source.
Therefore, if you and your handler are able to apply an extractor to your shared 100 digits,
the output will be 40 digits about which the adversary knows almost nothing.

\subsection{Removing Randomness Altogether}

We've seen that pseudorandom generators enable us to reduce the quantity of randomness needed
in computing, and randomness extractors enable us to reduce the quality of randomness needed
(as well as giving us new pseudorandom generators).
Can we remove the randomness altogether?

Although randomness is provably necessary for cryptography, the situation is unclear for algorithms.
In fact, for over twenty years the only efficient algorithm for testing primality was randomized.
Yet recently, several researchers managed to
give an efficient deterministic algorithm (an algorithm which uses no randomness) to test primality.
Could something similar be true for all randomized algorithms, i.e., there is a deterministic algorithm
which doesn't take too much longer than the corresponding randomized algorithm?

Many computer scientists believe the answer is yes:
we can remove all the randomness from algorithms and pay only a modest price.
However, proving this is well beyond the scope of current methods, and this problem is related to
the famous NP vs.\ P question.

Moreover, even this ``modest price'' to pay for a deterministic algorithm would likely be too high in practice.
For instance, the deterministic primality test is impractical, and that is unlikely to change anytime soon.
Randomness in algorithms seems with us to stay, and questions about the power of randomness remain
some of the most intriguing in computer science.

\section{Conclusions}

We've seen several examples of the power of randomness in computer science:
in various randomized algorithms and in cryptography.  In fact, randomness appears useful
in just about every area of computer science.  We've also discussed methods of
getting good randomness from a small amount of high-quality randomness and from a
large amount of low-quality randomness.

The topics we've discussed
involve interesting uses of probability and other areas of mathematics,
such as number theory.  Curiously, the famous British mathematician G.H.\ Hardy predicted that
number theory would never have any application.
Nevertheless, number theory is particularly useful in cryptography.

Using all of this mathematics, the results we discussed have been rigorously proved.
Rigorous proof is important, particularly in the area of cryptography.
When computer security is based only on the designers and their friends not being able to break a
proposed cryptosystem, the result is often an insecure system.
When the security is based on rigorous proof, this cannot happen.

To be fair, theorems in modern cryptography typically do make assumptions:
the theorems often state that if some problem, like factoring,
is hard, then the system is secure.
Yet having such clean assumptions was a breakthrough.
Since mathematicians have worked for
decades (or arguably centuries) on fast algorithms for factoring, all to no avail, it is unlikely
that a fast algorithm is around the corner.
(To be fair, if quantum computers can be built, then there are efficient algorithms to factor, so cryptographers
are working on new schemes.)

The role of randomness is just one topic in computer science where interesting mathematics
is used.
Theoretical computer science is the broad area of computer science focused on mathematical
aspects of computing.
Randomness and theoretical computer science are connected with many other areas of
math and science, which makes them such fascinating topics to study.

\section{Further Reading}

Other essays targeting a general audience can be found on the TheoryMatters website
\texttt{http://theorymatters.org/pmwiki/pmwiki.php?n=Main.SurveyCollection}.
For a longer, more mathematical introduction to randomness and computation,
emphasizing pseudorandomness and cryptography, see Oded Goldreich's essay \cite{Gol:rc-survey}.
Mitzenmacher and Upfal have an excellent introductory text on probability and computing
\cite{MitU} targeting advanced undergraduates.
At a similar level, there are high-quality introductory texts in
algorithm design \cite{KleT}, cryptography \cite{KatL}, and
the theory of computation \cite{Sip:book}.
Vadhan has an excellent article introducing pseudorandomness at the graduate level \cite{v:prg-survey}.  Finally,
Motwani and Raghavan have a thorough graduate-level text on randomized algorithms \cite{MotR}.

\section*{Acknowledgements}

I'm grateful to Juli Berwald, David Lieb, Sara Robinson,
Roland Stark, Gigi Taylor,
Daniel Zuckerman, and Paul~H. Zuckerman for very helpful comments.
This essay grew out of a talk first given at the Radcliffe Institute for Advanced Study.

\bibliographystyle{alpha}
\bibliography{/Users/diz/Documents/bibs/refs}

\end{document}